\documentclass[prb,twocolumn,superscriptaddress]{revtex4-2}

\usepackage{graphicx}
\usepackage{amsmath,amssymb}
\usepackage{natbib}

\usepackage[colorlinks,linkcolor=blue,anchorcolor=blue,urlcolor=blue,citecolor=blue]{hyperref}
\usepackage[all]{hypcap}

\usepackage{mathrsfs}
\usepackage{bm}
\usepackage{bbm}

\usepackage[utf8]{inputenc}
\usepackage[english]{babel}

\usepackage[T1]{fontenc}

\usepackage{newtxtext}
\usepackage{newtxmath}

\begin{document}
\title{Dynamical decoupling for realization of topological frequency conversion}
\author{Qianqian Chen}
\author{Haibin Liu}
\email{liuhb@hust.edu.cn}
\affiliation{School of Physics, International Joint Laboratory on Quantum Sensing and Quantum Metrology, Huazhong University of Science and Technology, Wuhan 430074, China}
\affiliation{Wuhan National Laboratory for Optoelectronics, and Huazhong University of Science and Technology, Wuhan 430074, China}
\affiliation{State Key Laboratory of Precision Spectroscopy, East China Normal University, Shanghai, 200062, China}
\author{Min Yu}
\author{Shaoliang Zhang}
\affiliation{School of Physics, International Joint Laboratory on Quantum Sensing and Quantum Metrology, Huazhong University of Science and Technology, Wuhan 430074, China}
\author{Jianming Cai}
\affiliation{School of Physics, International Joint Laboratory on Quantum Sensing and Quantum Metrology, Huazhong University of Science and Technology, Wuhan 430074, China}
\affiliation{Wuhan National Laboratory for Optoelectronics, and Huazhong University of Science and Technology, Wuhan 430074, China}
\affiliation{State Key Laboratory of Precision Spectroscopy, East China Normal University, Shanghai, 200062, China}
\date{\today}

\begin{abstract}
The features of topological physics can manifest in a variety of physical systems in distinct ways. Periodically driven systems, with the advantage of high flexibility and controllability, provide a versatile platform to simulate many topological phenomena and may lead to novel phenomena that can not be observed in the absence of driving. Here we investigate the influence of realistic experimental noise on the realization of a two-level system under a two-frequency drive that induces topologically nontrivial band structure in the two-dimensional Floquet space. We propose a dynamical decoupling scheme that sustains the topological phase transition overcoming the influence of dephasing. Therefore, the proposal would facilitate the observation of topological frequency conversion in the solid state spin system, e.g. NV center in diamond.
\end{abstract}

\maketitle

\section{Introduction}

Since the discovery of the quantized Hall effects \cite{klitzing1980,tsui1982,thouless1982,laughlin1981,Halperin1982,avron1983,laughlin1983}, topology is of abiding interest as a mechanism for generating extremely robust quantum mechanical phenomena that are insensitive to microscopic details. The finding that the Bloch bands of solid state systems can possess non-trivial topological characteristics has led to a whole new class of materials \cite{hasan2010,bernevig2013} which host a variety of remarkable phenomena and potential applications ranging from semiconductor spintronics \cite{vzutic2004} to topological quantum computation \cite{nayak2008}.
While many physical systems exhibit topological phenomena in condensed matters, the more controllable quantum simulator platforms keep emerging \cite{Martin2017,lin2016,ozawa2016,nathan2018,shu2018}, in which the spatial degrees of freedom can be replaced by the other kinds of degrees of freedom (e.g. optical or temporal degrees of freedom). This allows high controllability of topological phenomena and the increasing effective dimensionality of the system.
Topological phenomena induced by periodic driving \cite{Lindner2011,yao2007,oka2009,inoue2010,titum2015,Kitagawa2010,jotzu2014} can not only exhibit analogous behavior to those in real space but also demonstrate novel features as compared with static systems \cite{nathan2015,Rudner2013,Kitagawa2010,carpentier2015,kundu2013,jiang2011,titum2016}. For example, exotic topological phenomena such as anomalous Anderson-Floquet edge modes \cite{kundu2013,jiang2011,titum2016} that are described by topological invariants are unique periodically driven phases \cite{nathan2015,Rudner2013}, while the bulk Floquet bands are topologically trivial. New classifications of topological phases in terms of the spectra of Floquet operators were established \cite{Kitagawa2010,carpentier2015}. In particular, real-space lattice can be replaced by multidimensional Floquet lattice, resulting in temporal analogies of many-body phenomena in real-space \cite{shu2018,Martin2017,nathan2018}. One interesting example is the topological quantized pumping of photons between externally supplied circularly polarized electromagnetic modes, which represents a counterpart of the transverse current in a conventional topological insulator \cite{Martin2017}.

Recently, a synthetic quantum Hall effect with a two-tone drive has been studied experimentally using a single nitrogen-vacancy (NV) center in diamond \cite{boyers2020}. In this work, we investigate the effect of noise on the model of a two-level system under a two-frequency drive (with incommensurate frequencies $\omega_1$ and $\omega_2$ respectively) that induces topologically nontrivial band structure in the two-dimensional Floquet space  \cite{Martin2017}.
To be more specific, we carefully analyse the feasibility of realizing such a multidimensional Floquet lattice using a single electron spin in diamond under the influence of realistic noise. We demonstrate that the topological properties in this Floquet lattice will be affected by magnetic field fluctuation, namely the quantized pumping of energy deteriorates with increasing noise. We propose a dynamical decoupling strategy incorporating the two-frequency driving that allows to observe the phenomenon of topological frequency conversion in the solid-state spin system in diamond. Our results show that the solid-state spin system may be an appealing platform for the further investigation of topological phenomena induced by periodic driving.

\section{Floquet lattice and BHZ model}
We consider a time-dependent system with multiple-driven frequencies. The basis states of the system are denoted as $\left\vert {\alpha} \right\rangle$, which for example can represent the spin states. The system Hamiltonian is written as
\begin{equation}
  \label{eq:H_general}
  {\cal{H}}(t)=\sum_{{\alpha},{\beta}}H^{{\alpha}{\beta}}\left[ {\bm {\varphi}}(t) \right]\left| {\alpha}\rangle\langle {\beta} \right|
\end{equation}
with
\begin{equation}\label{}
  {\bm {\varphi}}(t)={\bm {\omega}} t+{\bm {\varphi}}_0,
\end{equation}
where ${\bm {\varphi}}=({\varphi}_1,{\varphi}_2,\dots)$, ${\bm {\varphi}}_0=({\varphi}_{01},{\varphi}_{02},\dots)$ and ${\bm {\omega}}=(\omega_1,\omega_2,\dots)$ represent the different frequencies of drives. The periodic Hamiltonian elements $H^{{\alpha}{\beta}}\left[ {\bm {\varphi}}(t) \right]$ can be expanded in terms of its Fourier components
\begin{equation}\label{}
  H^{{\alpha}{\beta}}\left[ {\bm {\varphi}}(t) \right]=\sum_{{\bf p}}h_{\bf p}^{{\alpha}{\beta}}e^{-i{\bf p}\cdot{\bm {\varphi}}(t)}.
\end{equation}
Based on Floquet theorem, the wave function can be expressed as
\begin{equation}\label{eq:wf}
  \left\vert \psi(t) \right\rangle=\sum_{{\alpha},n_1,n_2,\dots}\phi_{{\bf n}}^{\alpha} e^{-i(E+{\bf n}\cdot {\bm {\omega}}) t}\left\vert {\alpha} \right\rangle,
\end{equation}
where ${\bf n}=(n_1,n_2,\dots)$ denotes the indices of drives for different frequencies and $E$ is the energy of the system without drives. Substituting the wave function into the Schr{\"o}dinger equation results in the following tight-binding eigenvalue problem
\begin{equation}\label{eq:eigenproblem}
  \left( E+{\bf n}\cdot{\bm {\omega}} \right)\phi_{\bf n}^{\alpha}=\sum_{{\bf p},{\beta}}h_{{\bf p}}^{{\alpha}{\beta}}\phi_{{\bf n}-{\bf p}}^{\beta}.
\end{equation}
The term $n_i\omega_i$ in ${\bf n}\cdot{\bm {\omega}}=\sum_{i}n_i\omega_i$ has the physical meaning of the energy pumping arising from $n_i$ photons absorbed or emitted by the drive with the frequency $\omega_i$. Indeed, the term $e^{\pm i n_i \omega_it}$ in Eq.~\eqref{eq:wf} can be mapped into the picture of a Floquet lattice by
\begin{equation}\label{eq:intp}
    e^{\pm in_i\omega_it}\leftrightarrow \sum_\mathbf{n}c_{\mathbf{n}\mp n_i\hat{i}}^\dagger c_\mathbf{n},
\end{equation}
where $c^\dagger_{\mathbf{n}}(c_{\mathbf{n}}$) is the creation (annihilation) operator acting on the $\mathbf{n}=(n_1,n_2,\dots)$ site of the Floquet lattice, and $\hat{i}$ is the unit direction along the $i\textsuperscript{th}$ dimension. The term ${\bf n}\cdot{\bm {\omega}}$ describes the on-site energy of each lattice site, and $\omega_i$ is the potential-energy gradient over a lattice constant along the $i\textsuperscript{th}$ dimension.

%-------------------------------Figure 1-----------------------------------
\begin{figure}[t]
  \centering
  \includegraphics[width=1\columnwidth]{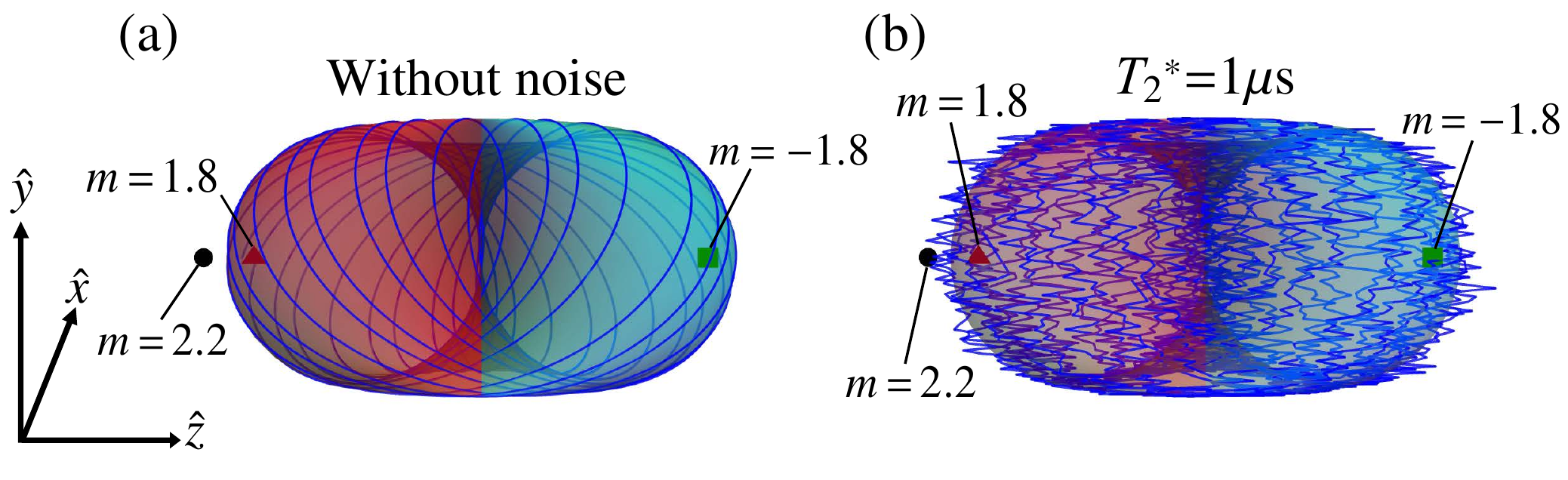}
  \caption{The trajectory (blue lines) of the terminal point of ${\bf h}(t)$ without noise ({\bf a}) and with the magnetic noise in the direction of $\hat{z}$ ({\bf b}), where the initial point of ${\bf h}(t)$ is always set at the origins of the coordinates. The standard deviation of the magnetic noise in ({\bf b}) is $\sigma_B= 1/(\sqrt{2}T^*_2)$ with $T_2^*=1\, \mu\rm{s}$. The closed and smooth manifold is formed by the trajectory of the terminal point of $\mathbf{h}$ (without noise) when $t\to\infty$, where the red (left half) part of the manifold has its norm vector outwards, the cyan (right half) part inwards, resulting in opposite signs of the Chern number in the topological region $-2<m<0$ and $0<m<2$. The circle, triangle (rectangle) denote $\mathbf{h}=0$ (where the Hamiltonian has band touching points) for the case of $m=2.2,\,1.8\,\,(-1.8)$, respectively, corresponding to topological trivial (outside the manifold) and nontrivial region (inside the manifold). Note that for clarity, the trajectories are set fixed with respect to different $m$, while the origins of the coordinate systems shift with different $m$ (see Eq.~\eqref{eq:h}). The coordinate system denoting the directions is on the left bottom. }
  \label{fig:traj}
\end{figure}
%-------------------------------Figure 1-----------------------------------

%
Following Ref.~\cite{Martin2017}, we consider the Hamiltonian of the BHZ model \cite{bernevig2006} with two-frequency drive as
\begin{equation}
  \begin{split}
    \frac{{\cal{H}}}{\eta}=\,&\sin{\left( {\omega}_1t+{\varphi}_{01} \right)}{\sigma}_x+\sin{\left( {\omega}_2t+{\varphi}_{02} \right)}{\sigma}_y\\
    &+\left[ m-\cos{\left( {\omega}_1t+{\varphi}_{01} \right)}-\cos{\left( {\omega}_2t+{\varphi}_{02} \right)} \right]{\sigma}_z\\
    &\equiv \left[ {\bf h}_1\left(\omega_1 t \right)+{\bf h}_2\left(\omega_2 t\right) \right]\cdot{\bm \sigma}
    \label{eq:m}
  \end{split}
\end{equation}
where $\omega_1$ and $\omega_2$ are two frequencies of the drive, ${\varphi}_{01}$ and ${\varphi}_{02}$ are the initial phase, $\eta$ defines the overall energy scale and ${\bm {\sigma}}=({\sigma}_x,{\sigma}_y,{\sigma}_z)$ are the Pauli matrices for the spin degree of freedom. And
\begin{subequations}\label{eq:h}
  \begin{align}
    {\bf h}_1(\omega_1 t)&\equiv\eta\left( \sin{\left( \omega_1 t+{\varphi}_{01} \right)},0,\frac{m}{2}-\cos{\left( \omega_1 t+{\varphi}_{01} \right)} \right)\\
    {\bf h}_2(\omega_2 t)&\equiv\eta\left( 0,\sin{\left( \omega_2 t+{\varphi}_{02} \right)},\frac{m}{2}-\cos{\left( \omega_2 t+{\varphi}_{02} \right)} \right).
  \end{align}
\end{subequations}
The vector ${\bf h}(\omega_1 t,\omega_2 t)\equiv{\bf h}_1(\omega_1 t)+{\bf h}_2(\omega_2 t)=h_x\hat{x}+h_y\hat{y}+h_z\hat{z}$ acts as a "Zeeman field" applied to a "pseudo-spin" ${\bm \sigma}$ of a two level system.
With incommensurate ratio between $\omega_1$ and $\omega_2$, the quasienergy operator $H$, expressed in the Floquet lattice representation by using Eq.~\eqref{eq:intp}, becomes
\begin{equation}\label{}
  \begin{split}
    H=&\eta\sum_{n_1,n_2}\bigg\{-\left( \frac{i}{2}{\sigma}_x+\frac{1}{2}{\sigma}_z \right)e^{i{\varphi}_{01}}c_{n_1-1,n_2}^{\dagger} c_{n_1,n_2}\\
    &-\left( \frac{i}{2}{\sigma}_y+\frac{1}{2}{\sigma}_z \right)e^{i{\varphi}_{02}}c_{n_1,n_2-1}^{\dagger} c_{n_1,n_2}
    +h.c.\bigg\}\\
    &+\sum_{n_1,n_2}\left(\eta m{\sigma}_z-n_1\omega_1-n_2\omega_2\right) c_{n_1,n_2}^{\dagger} c_{n_1,n_2},
  \end{split}
\end{equation}
After Fourier transformation of the hopping part by $c_{\mathbf{n}}=1/\sqrt{N}\sum_\mathbf{q}c_\mathbf{q}e^{i\mathbf{n}\cdot\mathbf{q}}$ and $c_{\mathbf{n}}^\dagger=1/\sqrt{N}\sum_\mathbf{q}c_\mathbf{q}^\dagger e^{-i\mathbf{n}\cdot\mathbf{q}}$, where $N$ is the number of Floquet lattice sites along one dimension, the quasienergy operator $H$ is of the form
\begin{equation}\label{}
  H=\sum_{q_1,q_2}{\cal{H}}_{q_1,q_2}c_{q_1,q_2}^{\dagger} c_{q_1,q_2}+\sum_{n_1,n_2}(-n_1\omega_1-n_2\omega_2)c_{n_1,n_2}^\dagger c_{n_1,n_2},
\end{equation}
where ${\cal{H}}_{q_1,q_2}$ takes the form
\begin{equation}\label{eq:Hq}
\begin{split}
{\cal{H}}_{q_1,q_2}=\,&\sin{\left( q_1+{\varphi}_{01} \right)}{\sigma}_x+\sin{\left( q_2+{\varphi}_{02} \right)}{\sigma}_y\\
    &+\left[ m-\cos{\left( q_1+{\varphi}_{01} \right)}-\cos{\left( q_2+{\varphi}_{02} \right)} \right]{\sigma}_z\\
\end{split}
\end{equation}
Because ${\bm {\omega}}$ can be interpreted as force, the momentum ${\bf q}$ evolves as $\mathbf{q}=(q_1,q_2)=(\omega_1 t,\omega_2 t)$. As a result, ${\cal{H}}_{q_1,q_2}$ has a period of $2\pi$ with respect to $q_i$ ($i=1,2$), which leads to an analogy with Brillouin zone: the Floquet zone (FZ). In the adiabatic limit $\eta\gg\max_i\{\omega_i\}$, the topological phase is characterized by Chern number
\begin{equation}
  C=\frac{1}{2\pi}\int_{\rm{FZ}}d{\bf q}\Omega_{\bf q},
  \label{eq:C}
\end{equation}
where the Berry curvature $\Omega_{\bf q}$ is defined as \cite{berry1984}
\begin{equation}\label{}
  \Omega_{\bf q}=i\left(\langle \partial_{q_1}\psi| \partial_{q_2}\psi \rangle-\langle \partial_{q_2}\psi| \partial_{q_1}\psi\rangle \right),
\end{equation}
and $\left|\psi\right >$ is the Bloch wave function of ${\cal{H}}_{q_1,q_2}$.
The Chern number of this system is \cite{Martin2017}
\begin{equation}
  C=\left\{
    \begin{aligned}
     -1 & \,\,\,\,\, &-2<m<0\\
      1 & \,\,\,\,\, & 0<m<2 \\
      0& \,\,\,\,\,  & \rm{otherwise}
    \end{aligned}\right.
\end{equation}
which yields topologically nontrivial band structure in the gap parameter region $\left\vert m \right\vert<2$ and the trivial region $\left\vert m \right\vert>2$.
The minimum gap $\Delta$ for this model is
\begin{equation}\label{}
  \Delta=\eta\min\left( \left\vert \left\vert m \right\vert-2 \right\vert,\left\vert m \right\vert \right).
\end{equation}
The two critical points where the degeneracy occurs are $m=0$ and $m=2$, where the system undergo topological phase transitions.

When the ratio between ${\omega}_i$ and ${\omega}_j$ ($i\ne j$) is incommensurate, the system is quasi-periodic. In the long time limit, the FZ is entirely sampled by the trajectory ${\bm {\varphi}}=\left( {\omega}_1 t+{\varphi}_{01},{\omega}_2 t+{\varphi}_{02} \right)$, which can also be reflected by the trajectory of the terminal point of ${\bf h}(t)$ as shown in Fig.~\ref{fig:traj}(a).
The total work $\left\langle{ E_k}\right\rangle$ done by the frequency mode $k$ is defined as
\begin{equation}
  \left\langle{ E_k}\right\rangle=\left\langle \psi(0) \right\vert\int_0^tdtU(t)^{\dagger}\partial_t{\left( {\bf h}_k\cdot{\bm \sigma} \right)}U(t)\left\vert \psi(0) \right\rangle,
  \label{eq:E}
\end{equation}
where the initial state $\left\vert\psi(0)\right\rangle\equiv\left\vert\psi(t=0)\right\rangle$ is one eigenstate of Hamiltonian Eq.~\eqref{eq:Hq} and $U(t)$ is the system's evolution operator. The energy pumping rate averaged over time is \cite{Martin2017}
\begin{equation}
  \begin{aligned}
    P_1=\frac{d\left\langle{ E_1}\right\rangle}{dt}={\omega}_1{\omega}_2{\sigma}_{xy} ={\omega}_1{\omega}_2\left(\frac{C}{2\pi}\right),
    \label{eq1}
  \end{aligned}
\end{equation}
where the Hall conductivity is ${\sigma}_{xy}=C/(2\pi)$. Here, the analogy has been made between the force ${\bm {\omega}}$ and the electric field  $\bm{\mathcal{E}}=(\mathcal{E}_x,\mathcal{E}_y)\equiv{\bm {\omega}}=(\omega_1,\omega_2)$ with the assumption that the effective charge is 1. One shall note that $P_2=d\left\langle{ E_2}\right\rangle/{dt}= - P_1$, which allows to extract the Chern number as $  C=\pi\left(P_1-P_2\right)/(\omega_1\omega_2)$ from the total work and the energy pumping rate.
\section{Realization of Floquet lattice with NV center}

In this section, we propose a scheme to realize the Floquet lattice and BHZ model using a solid-state spin system. The Hamiltonian Eq.~\eqref{eq:m} of the Floquet lattice can be realized by a highly controllable two-level solid-state system, which is a negatively charged NV center in diamond consisting of a substitutional nitrogen atom and an adjacent vacancy \cite{doherty2013,schirhagl2014,prawer2014}. The electrons around the defect form an effective electron spin with a spin-triplet ground state ($S = 1$).
In our experimental proposal, an external static magnetic field $B$ is applied to be parallel to the NV symmetry axis (also defined as $z$ axis), which enables both the NV electron spin and the host $\textsuperscript{14}$N nuclear spin to be polarized by optical excitation \cite{jacques2009,van2012}. The Hamiltonian of the electronic ground state of the NV center is
\begin{equation}\label{}
  {\cal{H}}_{\rm{NV}}=DS_z^2+\gamma B S_z,
\end{equation}
where $S_z$ is the angular momentum operator for spin-1, $D=(2\pi) 2870 \, \rm{MHz}$ is the zero-splitting, $\gamma=(2\pi) 2.8 \,\rm{MHz/G}$ is the gyromagnetic ratio of the NV electron, and $B$ is the external magnetic field along the NV axis.
The NV center spin is radiated by microwave (MW) driving fields with time-dependent amplitude and phase modulation (see Eq.~\eqref{eq:H_original}), the direction of which is perpendicular to the NV axis. Such microwave driving fields can be generated by arbitrary waveform generators (AWG) with a very high precision. The microwave driving fields selectively excite the transition between the electronic spin levels $\left\vert m_s=0 \right\rangle$ and $\left\vert m_s=-1 \right\rangle$, which represents the qubit used in the scheme. The applied magnetic field $B$ lifts the degeneracy of the spin sublevels $\left\vert m_s=\pm 1 \right\rangle$, thus the level $\left\vert m_s= 1 \right\rangle$ remains unaffected by the microwave driving fields due to a large detuning. The probability in $\left\vert m_s=0 \right\rangle$ can be read out via spin-dependent fluorescence detection during optical excitation. Under the driving of the engineered microwave fields, the Hamiltonian of the NV center spin can be written as
\begin{equation}
  \label{eq:H_original}
  \begin{split}
    {\cal{H}}_{\rm{NV}}=\frac{{\omega}_0}{2}{\sigma}_z&+2{\eta}\sin{\left( {\omega}_1 t+{\varphi}_{01} \right)}\cos{\left[ \left( {\omega}_0-{\omega}' \right)t \right]}{\sigma}_x\\
    &-2{\eta}\sin{\left( {\omega}_2t+{\varphi}_{02} \right)}\sin{\left[ \left( {\omega}_0-{\omega}' \right)t \right]}{\sigma}_x,
  \end{split}
\end{equation}
where $\omega_0$ is the energy level difference, $\sigma_x$ and $\sigma_z$ are the Pauli operators in the basis of $\{ \left\vert m_s=-1 \right\rangle, \left\vert m_s= 0 \right\rangle\}$. The amplitudes of the microwave fields are modulated with the frequencies $\omega_1$ and $\omega_2$ respectively. In additional, we introduce the following time-dependent phase as
\begin{equation}
  \begin{split}
    {\omega}'(t)=2{\eta} m&-\frac{2{\eta}}{{\omega}_1t}\left[ \sin{\left( {\omega}_1t+{\varphi}_{01} \right)}-\sin{{\varphi}_{01}} \right]\\
    &-\frac{2{\eta}}{{\omega}_2t}\left[ \sin{\left( {\omega}_2t+{\varphi}_{02} \right)}-\sin{{\varphi}_{02}} \right].
  \end{split}
\end{equation}
We remark that the manipulation of the NV center spin by microwave with time-dependent amplitude and phase modulation has been experimentally realized, see e.g. \cite{Farfurnik2017,yu2020,shu2018,yang2020}. To get the effective Hamiltonian, we introduce the following transformation $U_0$ as
\begin{equation}
  U_0=e^{i\int_0^tds {\cal{H}}_0(s)}\equiv e^{i{\sigma}_z{\theta}(t)},
\end{equation}
where
\begin{equation}
  {\cal{H}}_0=\frac{{\omega}_0}{2}{\sigma}_z-{\eta}\left[ m-\cos{\left( {\omega}_1t+{\varphi}_{01} \right)}-\cos{\left( {\omega}_2t+{\varphi}_{02} \right)} \right]{\sigma}_z
\end{equation}
and
\begin{equation}
  \begin{aligned}
    {\theta}(t)\equiv \,\,& \frac{{\omega}_0}{2}t-{\eta} mt \\
    &+\frac{{\eta}}{{\omega}_1}\left[ \sin{\left( {\omega}_1t+{\varphi}_{01} \right)}-\sin{\varphi}_{01} \right]\\
    &+\frac{{\eta}}{{\omega}_2}\left[ \sin{\left( {\omega}_2t+{\varphi}_{02} \right)}-\sin{\varphi}_{02} \right]\\
    =\,\,&\frac{1}{2}\left[ {\omega}_0-{\omega}'(t) \right]t.
  \end{aligned}
\end{equation}
\begin{figure*}[t]
  \centering
  \includegraphics[width=0.92\textwidth]{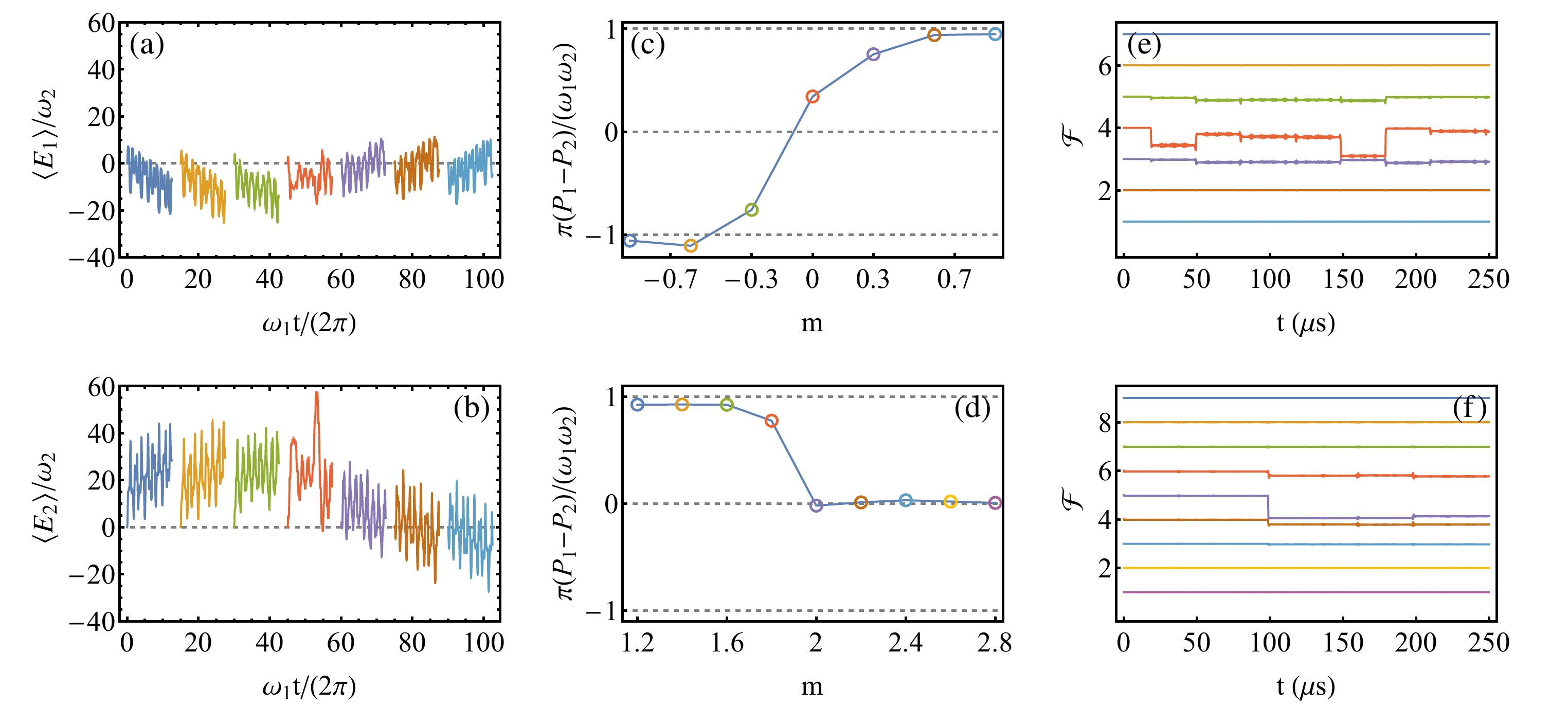}
  \caption{The total work $E_k$ and the fidelity for the driven spin system that is described by the Hamiltonian in Eq.~\eqref{eq:H_original}. ({\bf a}-{\bf b}) The numerical simulation of the total work performed by two frequency modes as a function of time for different values of the parameter $m$ (-0.9, -0.6, -0.3, 0.0, 0.3, 0.6, 0.9 from left to right). Each pair of lines is displaced on the time axis for clarity. ({\bf c}-{\bf d}) The energy pumping rates as a function of the parameter $m$ averaged up to $t=250\,\rm{\mu s}$. The quantized energy pumping is evident by the dashed gray lines that correspond to $dE/dt =C \omega_1\omega_2/(2\pi)$ with $C=0,\pm 1$. ({\bf e}-{\bf f}) The state fidelity as a function of time (see Eq.~\eqref{eq2}) between the evolving state from the initial eigenstate of ${\cal{H}}_{\rm{NV}}^{\rm{rot}}(0)$ and the corresponding instantaneous eigenstate of ${\cal{H}}_{\rm{NV}}^{\rm{rot}}(t)$ for the parameters across the phase transitions at $m=0$ ({\bf e}) and $m=2$ ({\bf f}). The values of $m$ are -0.9, -0.6, -0.3, 0.0, 0.3, 0.6, 0.9 from top to bottom in ({\bf e}), and 1.2, 1.4, 1.6, 1.8, 2.0, 2.2, 2.4, 2.6, 2.8 from top to bottom in ({\bf f}). We note that all fidelity-lines with different $m$ shall start at 1 when $t = 0$ and are offset vertically for clarity. We choose the amplitude of the microwave field as $2\eta=(2\pi)2 $ MHz, and the microwave modulation parameters are ${\omega}_1=(2\pi)50$ kHz and ${\omega}_2=(2\pi) 80.9$ kHz, the initial phases are ${\varphi}_{01}=\pi/10$ and ${\varphi}_{02}=0$, see Eq.~\eqref{eq:H_original}. }
  \label{fig:fig1}
  \label{fig:fidelity}
\end{figure*}
In the rotating frame as defined by $U_0$, and with the secular approximation, the effective Hamiltonian ${\cal{H}}_{\rm{NV}}^{\rm{rot}}(t)$ of the NV center spin (as driven by the microwave field given in Eq.~\eqref{eq:H_original}) can be written as follows
\begin{equation}
  \begin{split}
{\cal{H}}_{\rm{NV}}^{\rm{rot}}(t)
    =\,\,&\eta\{ \sin{\left( {\omega}_1t+{\varphi}_{01} \right)}{\sigma}_x+\sin{\left( {\omega}_2t+{\varphi}_{02} \right)}{\sigma}_y\\
    &+\left[ m-\cos{\left( {\omega}_1t+{\varphi}_{01} \right)}-\cos{\left( {\omega}_2t+{\varphi}_{02} \right)} \right]{\sigma}_z\},
    \label{eq:H_NV_rot}
  \end{split}
\end{equation}
which is identical with Eq.~\eqref{eq:Hq}. Therefore, we show that the scheme can simulate the Floquet lattice and BHZ model by engineering suitable microwave driving fields acing on the NV center spin.
Before we investigate the influence of noise, we perform numerical simulation using parameters from realistic experiments and demonstrate the feasibility of observing topological energy conversion in the present system. In the numerical simulation hereafter, we choose the amplitude of the microwave driving fields as $2\eta=(2\pi) 2$ MHz, and the frequencies of amplitude modulation are ${\omega}_1=(2\pi) 50$ kHz and ${\omega}_2=(2\pi) 80.9$ kHz. Thus, the ratio between $\omega_1$ and $\omega_2$ can be regarded as incommensurate, as long as the evolving time is far smaller than the common period of the two-drive (10 ms in our case). In additional, the initial phases are ${\varphi}_{01}=\pi/10$ and ${\varphi}_{02}=0$. We remark that these parameters are well achievable in experiments. The method in the numerical simulation hereafter is to construct the propagator governed by some time-dependent Hamiltonian $H(t)$ from the initial time to some final time, which can be approximated by $U(0,Ndt)\approx\prod_{n=1}^{N}\exp\left(-iH(ndt)dt\right)$. Considering the microwave field frequencies $\omega_1$ and $\omega_2$, we choose time discretization step as $dt=5$ns and assume the Hamiltonian is constant in this short time interval.
By numerical simulation, we calculate the total work $\langle E_1\rangle$ and $\langle E_2\rangle$ (see Eq.~\eqref{eq:E}) done by the frequency mode $\omega_1$ and $\omega_2$ respectively as a function of time. It can be seen from Fig.~\ref{fig:fig1} (a-b) that the feature of the total work changes qualitatively when the parameter crosses the critical point $m=0$. To make it more evident, we extract the slope of the total work in Fig.~\ref{fig:fig1} (a-b) by the method of linear regression and estimate the value of the energy pumping rate $P_1$ and $P_2$ (see Eq.~\eqref{eq1}). The results are shown in Fig.~\ref{fig:fig1} (c), which demonstrates that the pumping of energy between the driving modes with frequencies $\omega_1$ and $\omega_2$, averaged over time, occurs at a quantized rate. In particular, one can see that when the parameter $m$ changes from $m=-0.9$ to $m=0.9$, which goes across the critical parameter of the phase transition $m=0$, the direction of the energy conversion between the first and second frequency mode inverses, i.e. $\omega_1 \rightarrow \omega_2$ ($m<0$) and $\omega_1 \leftarrow \omega_2$ ($m>0$). The transition at the point $m=0$ corresponds to the phase transition with the topological invariant Chern number (see Eq.~\eqref{eq:C}) changing from $-1$ to $1$. In a similar way, we perform numerical simulation for the parameter $m\in[1.2,2.8]$. The result in Fig.~\ref{fig:fig1} (d) also shows the change of the quantized Chern number from 1 to 0 when $m$ passes the critical point at $m=2$.
The phase transition is caused by the closing of the energy gap, which can be reflected by the fidelity between the evolving state $\left\vert \psi(t) \right\rangle=U(t)\left\vert \psi_i(0) \right\rangle$, where the initial state $\left\vert \psi_i(0) \right\rangle$ is chosen to be the $i\textsuperscript{th}$ eigenstate of ${\cal{H}}_{\rm{NV}}^{\rm{rot}}(0)$, and the $i\textsuperscript{th}$ instantaneous eigenstate $\left\vert \psi_i(t) \right\rangle$ of ${\cal{H}}_{\rm{NV}}^{\rm{rot}}(t)$. The fidelity is defined as follows
\begin{equation}
  \begin{aligned}
    {\cal F}=&\left\langle \psi(t) \right\vert P_t\left\vert \psi(t) \right\rangle,
    \label{eq2}
  \end{aligned}
\end{equation}
where $P_t=\left| \psi_i(t)\rangle\langle \psi_i(t) \right|$. The fidelities around the critical point $m=0$ and $m=2$ are shown in Fig.~\ref{fig:fidelity} (e) and (f), respectively, where the closing of the gap leads to the prominent deterioration of the fidelity. In contrast, the fidelity is robust as long as ${\Delta}\gg \max_i \left\{ {\omega}_i  \right\}$. As an example, it can be seen that the fidelity remains near perfect throughout the evolution for the parameter ranging from $m=1.2$ to $m = 1.6$. When $m=1.8$, the energy gap ${\Delta}={\eta}\left\vert \left\vert m \right\vert-2 \right\vert=1.26\,\rm{MHz}$ is comparable to ${\omega}_2 =0.51\,\rm{MHz}$, and the quasi-adiabatic condition breaks down. For larger values of $m$, the gap reopens and the fidelity becomes robust again at $m=2.4$, which corresponds to ${\Delta}={\eta}\left\vert \left\vert m \right\vert-2 \right\vert=2.52\,\rm{MHz}> {\omega}_2=0.51\,\rm{MHz}$. Similar behaviors can be observed when the parameter $m$ changes from $m=-0.9$ to $m=0.9$, across the critical point $m=0$, see Fig.~\ref{fig:fidelity} (e). We remark that in the NV spin system the initial state $\left\vert \psi_i(0) \right\rangle$ can be prepared through Rabi oscillation, and the fidelity can be determined by the spin-dependent fluorescence measurement after a suitable microwave pulse \cite{yu2020}.

\section{Effect of noise and dynamical decoupling strategy}

For the NV center spin system, due to a large energy mismatch between the spin bath and the NV center spin, the dominant noise is dephasing caused by the longitudinal magnetic field fluctuation. In this section, we will show that this dephasing noise will affect the pumping quantization and make the topological phase transition much less prominent. Taking into account the influence of the dephasing noise, the system can be described by the following Hamiltonian $H_{\rm{NV}}^{\rm{rot}}$ as
\begin{equation}
  \begin{split}
    {\cal{H}}_{\rm{noise}}^{\rm{rot}}(t)=\,\,&\eta\big\{\sin{\left( {\omega}_1t+{\varphi}_{01} \right)}{\sigma}_x+\sin{\left( {\omega}_2t+{\varphi}_{02} \right)}{\sigma}_y\\
    &+\left[ m-\cos{\left( {\omega}_1t+{\varphi}_{01} \right)}-\cos{\left( {\omega}_2t+{\varphi}_{02} \right)} \right]{\sigma}_z\big\}\\
    &+{\delta} (t){\sigma}_z,
    \label{Eq-H-n}
  \end{split}
\end{equation}
where ${\delta}(t)$ represents the longitudinal magnetic noise.

For an NV center spin in an electron spin bath, the noise can be well described by the widely used classical Gaussian noise: the Ornstein--Uhlenbeck (OU) noise \cite{Dobrovitski2009, Witzel2012, Lange2010, Hanson2008, Wang2013}. For a nuclear spin bath, the OU noise is also applicable when the hyperfine interaction is anisotropic and the magnetic field is intermediate \cite{WenYang2016}. Thus for typical experiments in NV centers, we can simulate the noise $\delta(t)$ with the following iterative formula for OU process \cite{Dan96}
\begin{equation}
    \delta(t)=\delta(t-dt)e^{-\frac{dt}{\tau}}+y\sqrt{v\left(1-e^{-2\frac{dt}{\tau}}\right)},
    \label{eq:random}
\end{equation}
where $v$ is the variance of the magnetic noise, $y$ is a random variable (which is updated in every iteration) of the standard normal distribution,  and $\tau$ is the noise correlation time. We choose $\delta(t=0)$ as a random value of normal distribution with an average $0$ and a variance $v$, and the discretization step $dt$ is set as $dt=5 \,\rm{n s}$ in the simulation. The relation between the variance $v$ and the coherence time $T_2^*$ is $T_2^*=1/\sqrt{2 v}$.
\begin{figure}[t]
\includegraphics[width=1\columnwidth]{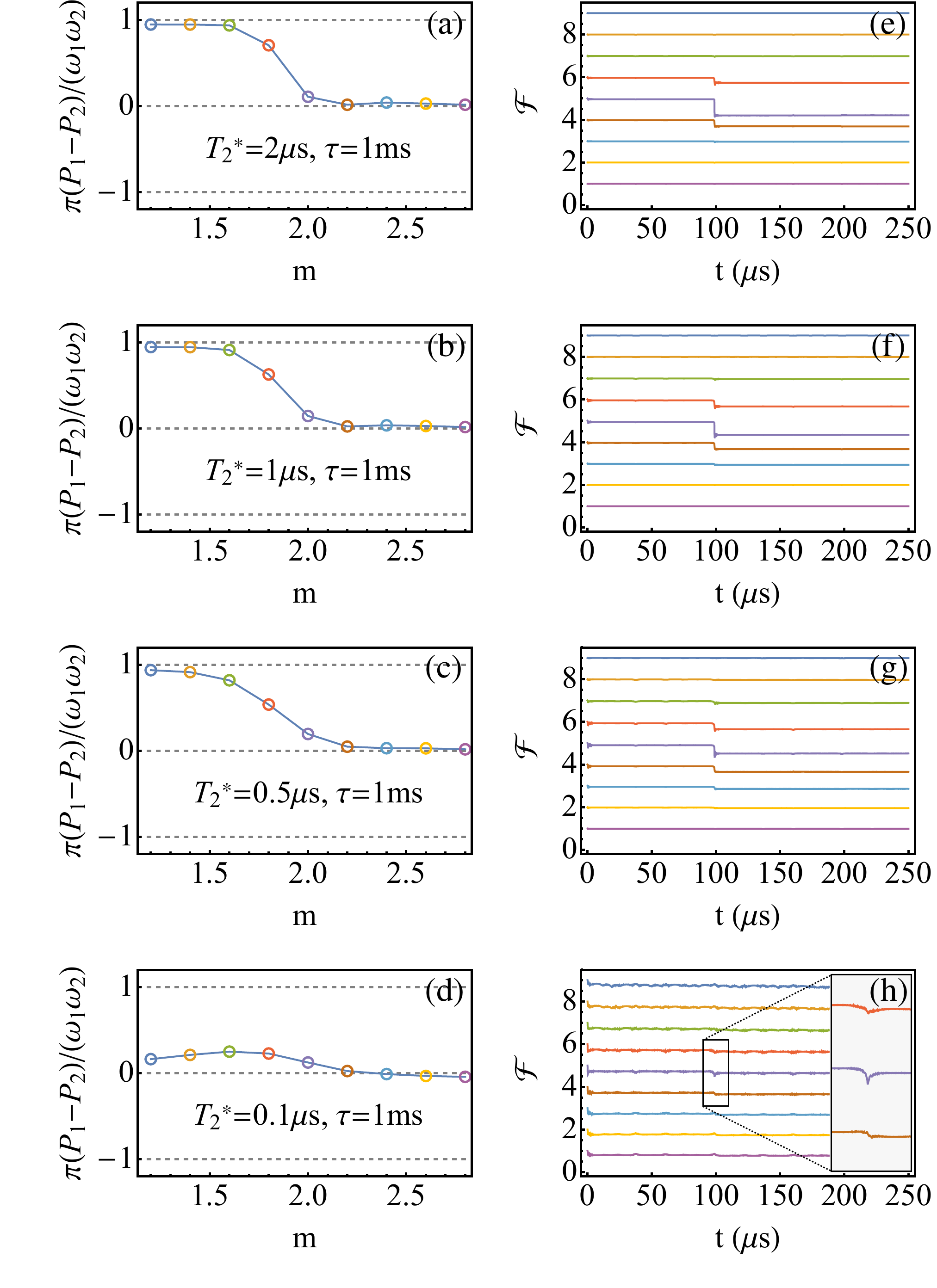}
\caption{
The energy pumping rate and the fidelity averaged over 1000 random instances.
({\bf a}-{\bf d}) The energy pumping rate as a function of the parameter $m$ for the NV center spin with the coherence time $T_2^*=2\,\rm{\mu s}$ ({\bf a}), $T_2^*=1\,\rm{\mu s}$ ({\bf b}), $T_2^*=0.5\,\rm{\mu s}$ ({\bf c}) and $T_2^*=0.1\,\rm{\mu s}$ ({\bf d}). The quantized pumping is marked by the dashed gray lines that correspond to $dE/dt =C \omega_1\omega_2/(2\pi)$ with $C=0,\pm 1$. ({\bf e}-{\bf h}) The state fidelity, between the evolving state from the initial eigenstate of ${\cal{H}}_{\rm{NV}}^{\rm{rot}}(0)$ and the corresponding instantaneous eigenstate of ${\cal{H}}_{\rm{NV}}^{\rm{rot}}(t)$, as a function of time (see Eq.~\eqref{eq2}). The NV center spin's coherence time is $T_2^*=2\,\rm{\mu s}$ ({\bf e}), $T_2^*=1\,\rm{\mu s}$ ({\bf f}), $T_2^*=0.5\,\rm{\mu s}$ ({\bf g}) and $T_2^*=0.1\,\rm{\mu s}$ ({\bf h}). The values of $m$ are 1.2, 1.4, 1.6, 1.8, 2.0, 2.2, 2.4, 2.6, 2.8 from top to bottom. We note that all fidelity-lines with different $m$ shall start at 1 when $t = 0$ and are offset vertically for clarity. The noise correlation time is $\tau=1\,\rm{ms}$. The inset of ({\bf h}) shows a magnified plot in the region where fidelities for $m=1.8$, $2.0$, $2.2$ should drop but are deteriorated by noise. }
\label{fig:noise}
\end{figure}
We calculate the pumping rate and the fidelity incorporating the influence of the spin-bath caused magnetic noise, the results obtained by averaging 1000 random instances are shown in Fig.~\ref{fig:noise}. It can be seen that the topological phase transition characterized by the energy pumping rate, see Fig.~\ref{fig:noise} (a-d), becomes less steep with the decreasing coherence time $T_2^*$ of the NV center spin (i.e. the increasing of the magnetic noise). In the meantime, the fidelity between the system's evolving state and the instantaneous eigenstate of ${\cal{H}}_{\rm{NV}}^{\rm{rot}}(t)$ deteriorates significantly, see Fig.~\ref{fig:noise} (e-h). In particular, when the standard deviation of the magnetic noise $\sigma_B$ reaches $(2\pi)1.125\,\rm{MHz}$ (corresponding to the NV center spin coherence time $T_2^*=0.1\,\rm{\mu s}$), see Fig.~\ref{fig:noise} (d) \& (h),  it becomes very hard to observe the critical phenomenon of the topological phase transition anymore. The typical feature of the close of the gap around $m=2$ disappears in the fidelity, as shown by Fig.~\ref{fig:noise} (h), which indicates that one can observe the deterioration caused by the noise through NV. Indeed, as is indicated in Fig.~\ref{fig:traj}(b), the magnetic noise blurs the profile of the otherwise closed and smooth surface that ${\bf h}(t)$ samples in the long time limit.
\begin{figure}[t]
  \centering
  \includegraphics[width=0.86\columnwidth]{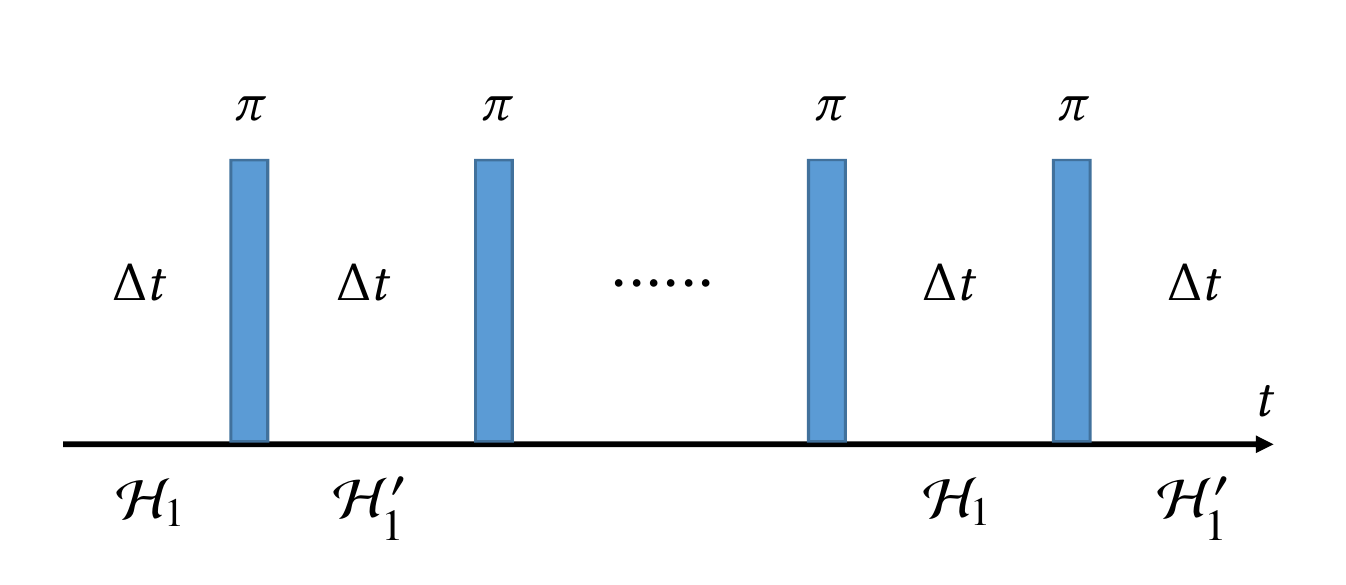}
  \caption{The dynamical decoupling sequence that consists of repetitive equally distant $\pi$ pulses with the inter-pulse Hamiltonian $\mathcal{H}_1$ and $\mathcal{H}_1'$, see Eq.~(\ref{Eq-H-n}, \ref{eq:Eff_Ham_2}). The scheme implements the effective Hamiltonian (Eq.~\eqref{eq:Eff_Ham}) for the observation of topological energy conversion under the influence of noise.}
  \label{fig:fig5}
\end{figure}

\begin{figure}[t]
%  \hspace{-0.8cm}
  \includegraphics[width=1\columnwidth]{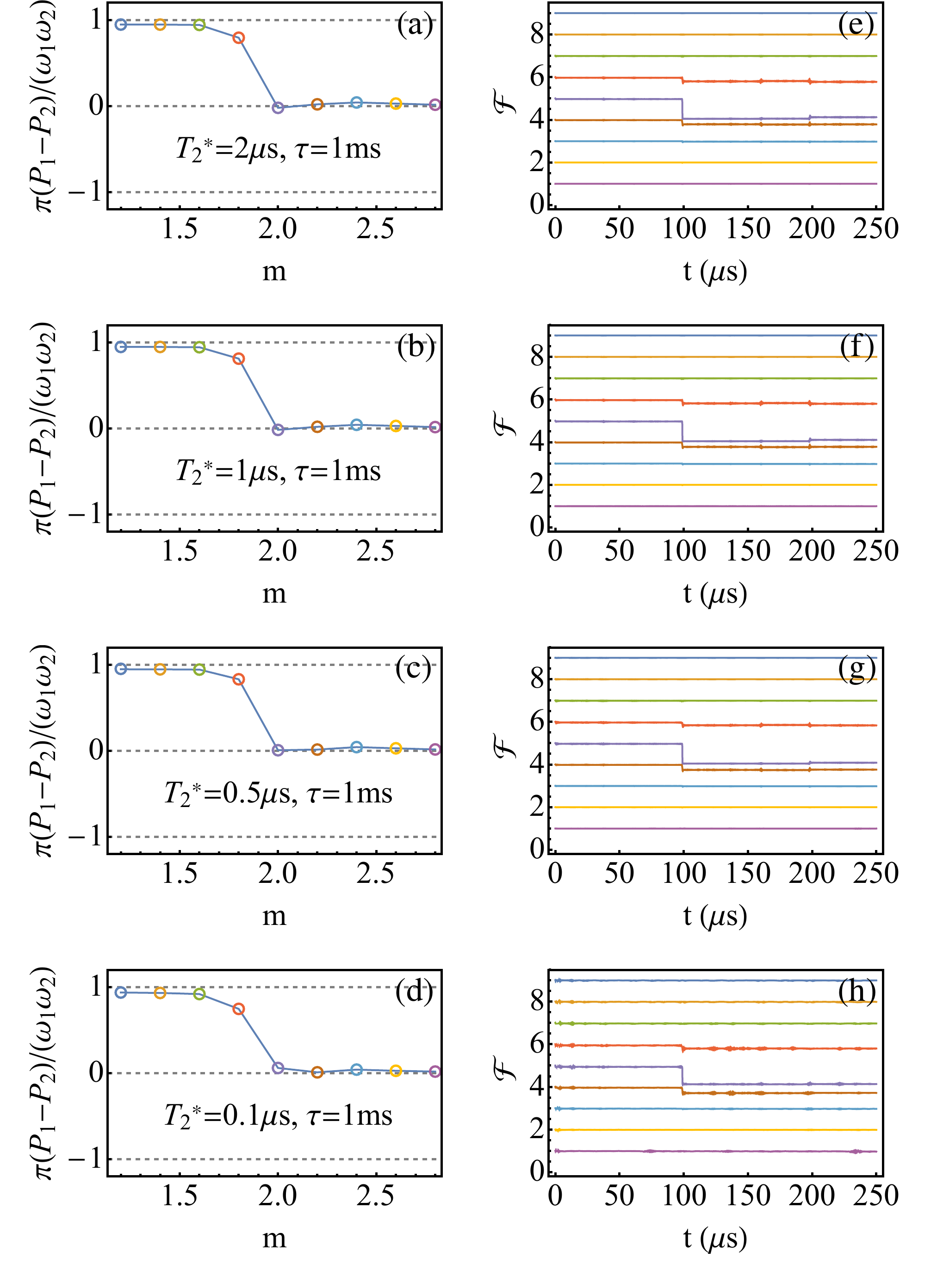}
  \caption{The energy pumping rate and the fidelity averaged over 1000 random instances with the dynamical decoupling pulses applied to restore the topological phase transition features. ({\bf a}-{\bf d}) The energy pumping rate as a function of the parameter $m$ for the NV center spin with the coherence time $T_2^*=2\,\rm{\mu s}$ ({\bf a}), $T_2^*=1\,\rm{\mu s}$ ({\bf b}), $T_2^*=0.5\,\rm{\mu s}$ ({\bf c}) and $T_2^*=0.1\,\rm{\mu s}$ ({\bf d}). The quantized pumping is marked by the dashed gray lines at $dE/dt =C \omega_1\omega_2/(2\pi)$ with $C=0,\pm 1$. ({\bf e}-{\bf h}) The state fidelity, between the evolving state from the initial eigenstate of ${\cal{H}}_{\rm{NV}}^{\rm{rot}}(0)$ and the corresponding instantaneous eigenstate of ${\cal{H}}_{\rm{NV}}^{\rm{rot}}(t)$, as a function of time (see Eq.~\eqref{eq2}). The NV center spin's coherence time is $T_2^*=2\,\rm{\mu s}$ ({\bf e}), $T_2^*=1\,\rm{\mu s}$ ({\bf f}), $T_2^*=0.5\,\rm{\mu s}$ ({\bf g}) and $T_2^*=0.1\,\rm{\mu s}$ ({\bf h}). The values of $m$ are 1.2, 1.4, 1.6, 1.8, 2.0, 2.2, 2.4, 2.6, 2.8 from top to bottom. We note that all fidelity-lines with different $m$ shall start at 1 when $t = 0$ and are offset vertically for clarity.  The noise correlation time is $\tau=1\,\rm{ms}$, and the inter-pulse time period $\Delta t$ is set to be 50 $\rm{n s}$.
  }
\label{fig:ddp}
\end{figure}
To overcome the constraint imposed by the noise and facilitates the observation of topological phase transition under the influence of noise, we propose a dynamical decoupling based scheme to counteract the deterioration caused by the noise and to observe topological energy conversion. We remark that various techniques of dynamical decoupling have been developed to prolong the coherence time of NV center in diamond, see e.g. \cite{Naydenov2011,cai2012}.
In our scheme, the OU Gaussian noise possess a Lorentzian spectrum which is soft-cutoff. According to \cite{Cywi2008,Pasini2010}, periodic pulses similar to Carr-Purcell-Meiboom-Gill (CPMG) ones are ideal for decoupling from an environment with a soft cut-off noise.
Here, we adopt the CPMG dynamical decoupling sequences incorporating with effective Hamiltonian engineering in order to observe topological energy conversion using the NV center spin system in diamond. The scheme is shown in Fig.~\ref{fig:fig5}, in which we propose to apply equally distant  $\pi$-pulses.
For each cycle $t\in [2(k-1)\Delta t,  2 k \Delta t]$ ($k$ is a positive integer), the system's Hamiltonian can be chosen as follows
\begin{equation}
    {\cal{H}}(t)=
  \begin{cases}
    {\cal{H}}_1(t)=
    {\cal{H}}_{\rm{noise}}^{\rm{rot}}(t) , & 2 (k-1)\Delta t < t \le (2 k-1) \Delta t \\
    {\cal{H}}'_1(t), &  (2k-1)\Delta t<t \le 2k \Delta t,
  \end{cases}
\end{equation}
where
\begin{equation}
  \begin{split}
    {\cal{H}}_1'(t)=\,\,&\eta\big\{\sin{\left( {\omega}_1t+{\varphi}_{01} \right)}{\sigma}_x-\sin{\left( {\omega}_2t+{\varphi}_{02} \right)}{\sigma}_y\\
    &-\left[ m-\cos{\left( {\omega}_1t+{\varphi}_{01} \right)}-\cos{\left( {\omega}_2t+{\varphi}_{02} \right)} \right]{\sigma}_z\big\}\\
    &+{\delta} (t){\sigma}_z.
  \end{split}
\end{equation}
We note that the implementation of the Hamiltonian $ {\cal{H}}_1(t)$ is shown in Eq.~\eqref{eq:H_original}. The Hamiltonian $ {\cal{H}}_1'(t)$ can be realized by the following engineered microwave fields in the direction perpendicular to the NV axis
\begin{equation}
  \begin{split}
    f(t)=\,\, &2{\eta}\sin{\left( {\omega}_1 t+{\varphi}_{01} \right)}\cos{\left[ \left( {\omega}_0-{\omega}' \right)t \right]}\\
    &+2{\eta}\sin{\left( {\omega}_2t+{\varphi}_{02} \right)}\sin{\left[ \left( {\omega}_0-{\omega}' \right)t \right]} .
\end{split}
\end{equation}
Thus, under the instantaneous $\pi$-pulse limit we can get the effective Hamiltonian ($(2k-1)\Delta t<t \le 2k \Delta t$) as
\begin{equation}\label{}
  \begin{split}
    {\cal{H}}_2(t)\equiv\,\,&   \sigma_x {\cal{H}}_1'(t)\sigma_x\\
    =\,\,&\eta\big\{\sin{\left( {\omega}_1t+{\varphi}_{01} \right)}{\sigma}_x+\sin{\left( {\omega}_2t+{\varphi}_{02} \right)}{\sigma}_y\\
    &+\left[ m-\cos{\left( {\omega}_1t+{\varphi}_{01} \right)}-\cos{\left( {\omega}_2t+{\varphi}_{02} \right)} \right]{\sigma}_z\big\}\\
    &-{\delta} (t){\sigma}_z.  \label{eq:Eff_Ham_2}
  \end{split}
\end{equation}

\begin{figure}[t]
  \includegraphics[width=1\columnwidth]{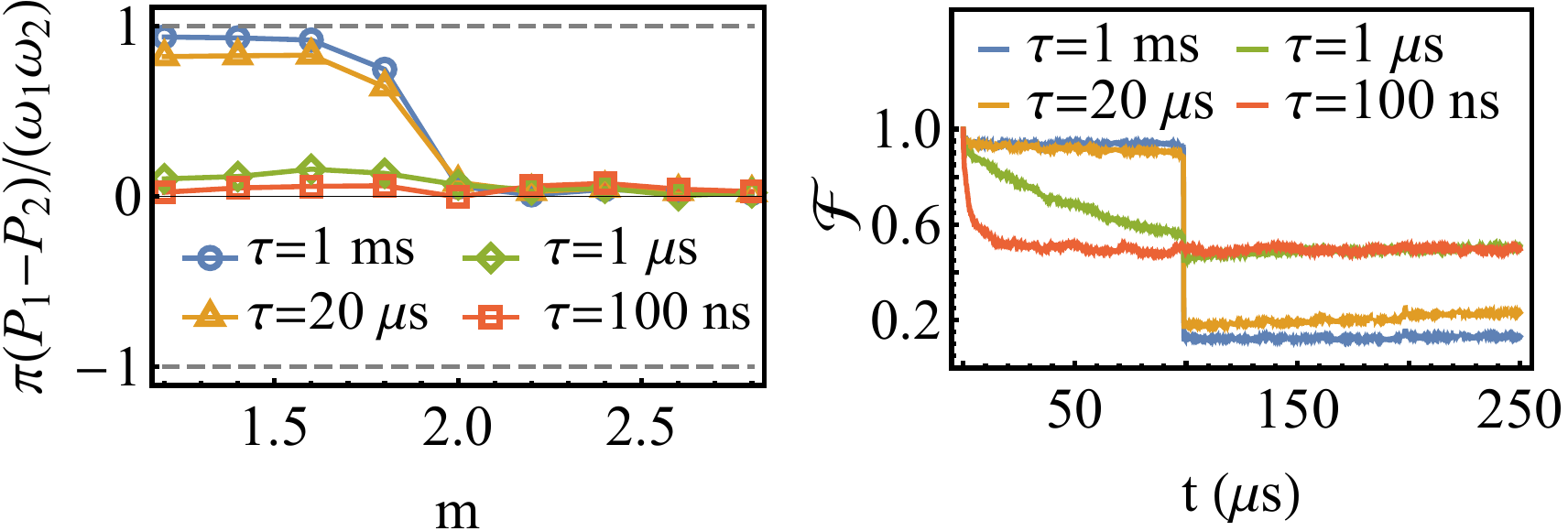}
  \caption{The energy pumping rates and fidelities with the dynamical decoupling pulses under a dephasing noise of $T_2^*=0.1\,\mu s$ and different correlation time. ({\bf a}) The energy pumping rate as a function of the parameter $m$ for the NV center spin with the noise of correlation time $\tau=1$ ms (the blue line), $\tau=20\,\mu$s (the orange line), $\tau=1\,\mu$s (the green line), and $\tau=100$ ns (the red line).
  ({\bf b}) The fidelities with $m=2$ and the noise of correlation time $\tau=1$ ms (the blue line), $\tau=20\,\mu$s (the orange line), $\tau=1\,\mu$s (the green line), and $\tau=100$ ns (the red line).
  }
\label{fig:ct}
\end{figure}
If the time period $\Delta t$ is much short compared with the oscillation of ${\cal{H}}_1(t)$ and ${\cal{H}}'_1(t)$, we can treat ${\cal{H}}(t)$ as constant during the time period $\Delta t$. In additional, we assume that the noise $\delta(t)$ remains almost the same in the first time period $\Delta t$ and the subsequent time period $\Delta t$, which is valid when the correlation time of the noise is much longer than $\Delta t$. Up to the first order of $\Delta t$, the effective Hamiltonian of the $2\Delta t$ period can be approximated as
\begin{equation}
   {\cal{H}}_{\rm{eff}}(t)\equiv\frac{{\cal{H}}_1(t)+{\cal{H}}_2(t)}{2}={\cal{H}}_{\rm{NV}}^{\rm{rot}}(t), \label{eq:Eff_Ham}
\end{equation}
which implies that the effect of noise can be mitigated. This can be verified explicitly. The system's evolution operator during the time period of $2\Delta t$ is
\begin{equation}
\label{eq:Heff}
  \begin{split}
    &\sigma_x \mathcal{T}e^{-i\int_{t+\Delta t}^{t+2\Delta t}{\cal{H}}_1'(\tau) d\tau }\sigma_x \mathcal{T}e^{-i\int_{t}^{t+\Delta t}{\cal{H}}_1(\tau)d\tau}\\
    =&\, \mathcal{T}e^{-i\int_{t+\Delta t}^{t+2\Delta t} {\cal{H}}_2(\tau) d\tau}\mathcal{T}e^{-i\int_{t}^{t+\Delta t}{\cal{H}}_1(\tau)d\tau} \\
    \approx &\, e^{-i\int_{t}^{t+2\Delta t}\left( {\cal{H}}_1(\tau)+{\cal{H}}_2(\tau) \right)/2d\tau}\\
    \equiv&\,e^{-i\int_{t}^{t+2\Delta t}{\cal{H}}_{\rm{eff}}(\tau)d\tau}
  \end{split}
\end{equation}
where $\mathcal{T}$ is the time-ordering operator, and the approximation of Eq.~\eqref{eq:Heff} is made to the first order of $\Delta t$ for Magnus expansion. Fig.~\ref{fig:ddp} shows the energy pumping rate and the fidelity based on the proposed dynamical decoupling strategy. It can be seen that the influence of the noise on the topological features is greatly suppressed. In particular, the scheme allows us to observe the quantized energy pumping rate and the topological phase transitions unambiguously, even the NV center spin's coherence time is as short as $T_2^*=0.1\,\rm{\mu s}$. The correlation time of the spin bath may vary over different NV center samples. The simulation results for different correlation time of noise is shown in Fig.~\ref{fig:ct}, which demonstrates that our proposal can mitigate noise to a great extent for typical correlation time of nuclear spin bath noise $\approx1$ ms \cite{WenYang2016} and electron spin bath noise $\approx20\,\mu$s \cite{Lange2010}.

\section{Conclusion and Discussion}
To conclude, in this work, we propose to realize the two-dimensional Floquet lattice using a single spin in diamond with a two-frequency drive at room-temperature. We demonstrate that the noise will affect the topological phenomena that the Floquet lattice exhibits. To overcome such a challenge imposed by the noise, we propose a strategy incorporating pulsed dynamical decoupling to observe topological energy conversion under the influence of noise. With detailed numerical simulation using realistic experimental parameters, we show the feasibility of the scheme using the NV center spin system. The work suggests that the NV center spin system (assisted by dynamical decoupling) may provide a versatile platform to realize the counterpart of real-space lattice, i.e., the Floquet lattice. With incommensurate frequencies, it is possible to further explore the quasicrystals with infinite lattice \cite{kalugin1985,levine1984,shechtman1984}. With commensurate frequencies, the Floquet lattice has the periodic boundary conditions such that the lattice contracts to a cylinder \cite{Martin2017}. Furthermore, one can apply multi-frequency drives to realize the multidimensional Floquet lattice that is not easy to construct in real space.

\section{Acknowledgements}
We acknowledge useful discussion about OU noises with Jiazhao Tian. This work is supported by the Young Scientists Fund of the National Natural Science Foundation of China (Grant No. 11804110), the National Natural Science Foundation of China (Grants No. 11874024, 11690032), the Open Project Program of Wuhan National Laboratory for Optoelectronics (Grant No. 2019WNLOKF002).

\bibliography{literature}

\end{document}